**Overcoming Van der Waals Forces in reconfigurable nanostructures**

Wang Zhang[1], Hao Wang[1*], Alvin T. L. Tan[1], Anupama Sargur Ranganath[1], Biao Zhang[2], Hongtao Wang[1], John You En Chan[1], Qifeng Ruan[1], Hailong Liu[3], Son Tung Ha[3], Dong Wang[4], Venkat K. Ravikumar[5], Hong Yee Low[1], Joel K.W. Yang[1,3*]

[1]Engineering Product Development, Singapore University of Technology and Design, Singapore 487372, Singapore. [2]Frontiers Science Center for Flexible Electronics, Xi'an Institute of Flexible Electronics (IFE) and Xi'an Institute of Biomedical Materials & Engineering (IBME), Northwestern Polytechnical University, 127 West Youyi Road, Xi'an 710072, China. [3]Institute of Materials Research and Engineering, Singapore 138634, Singapore. [4]Robotics Institute and State Key Laboratory of Mechanical System and Vibration, School of Mechanical Engineering, Shanghai Jiao Tong University, Shanghai 200240, PR China. [5]Advanced Micro Devices Singapore Pte Ltd, Singapore.

*Email: whchn@live.cn; joel_yang@sutd.edu.sg

***Abstract***

Reconfigurable metamaterials[1-3] require constituent nanostructures to demonstrate switching of shapes with external stimuli. For generality, such nanostructures would touch and stick to other surfaces in one of its configurations. Yet, a longstanding challenge is in overcoming this stiction caused by Van der Waals forces, which impedes shape recovery. Here, we introduce a stiff yet self-recovering material system based on acrylic acid, and tested it in high-aspect ratio structures, where recovery is weak. This designer material has a storage modulus of ~5.2 GPa at room temperature and ~90 MPa in the rubbery state at 150 °C, an order of magnitude higher than previous reports. A high-resolution resin for two-photon lithography was developed based on this polymer system, enabling 3D printing of nanopillars with diameters of ~400 nm and aspect ratio as high as ~10. Experimentally, we observed self-recovery as collapsed and touching structures overcome stiction to stand back up. We developed a theoretical model to explain the recoverability of these sub-micron structures. Reconfigurable structural colour prints and holograms were demonstrated, indicating potential applications of the material system as a shape memory polymer suitable for sub-micron reconfigurable metamaterials.

Reconfigurable structures find use in numerous areas, such as biological engineering[4], soft robotics[5], and flexible electronics[6]. At the macro scale, configuration transformation is feasible through exposure of stimuli-responsive materials to heat[7], water[8], light[9], electricity[10] and magnetic fields[11]. Pushing these reconfigurable structures to the nanoscale would enable new applications in nano optics[12], sensors[13] and micro robotics[14]. However, a longstanding challenge in reconfigurable nanostructures is related to their significantly higher surface to volume ratio where Van der Waals (VDW) forces[15] dominate. As VDW forces exceed the internal restoring forces as structures decrease in size, nanostructures that come in physical contact tend to "stick" and never detach[1,16].

Avoiding permanent deformation generally involves increasing the restoring forces and/or decreasing VDW forces. For instance, restoring forces can be increased through over-constrained nanostructure geometries[2,3], while VDW forces can be decreased by operating in a vacuum chamber[17] and/or surface treatments to reduce surface energy[18]. However, these strategies would not work for free-standing nanostructures as the reported stimuli materials are designed to be soft and flexible. Possessing low storage modulus of less than 10 MPa in the rubbery state, these structures will suffer from lateral collapse[19] or are too soft to overcome the VDW forces[15] during the attempted shape recovery (reset) process. Conversely, highly crosslinked rigid materials e.g. the proprietary IP-Dip resin[20] are unsuitable as they would fracture when deformed during the programming (set) process. Hence, new material systems are needed with sufficient flexibility to be deformed and collapse when set, yet possess strong capacity for self-recovery to overcome the influence of VDW forces and detach during reset. Simultaneously, suitable fabrication methods need to be developed to explore new applications of stimuli responsive materials at the nanoscale.

In this work, we investigated the reconfigurability of high aspect ratio (tall) nanopillars. This geometry provides little structural rigidity, thus allowing us to examine the self-recovery of the proposed material system. Concomitantly, these structures exhibit structural colours[21-24] that provide a convenient visual evidence for nanostructure recovery. We introduced acrylic acid-based shape memory polymers (SMPs) and developed a resin for two-photon polymerization lithography (TPL). We fabricated nanopillars with ~400 nm diameter and aspect ratio of ~10 by TPL. With the material developed, these tall and narrow structures can be deformed and stay in a collapsed and curled up configuration, yet remarkably overcome the VDW forces and recover when triggered by heat to achieve nanoscale reconfigurable structures.

High resolution reconfigurable structural colour print and hologram are realized to demonstrate a potential application of the proposed materials.

The design principle is shown schematically in Fig. 1aI. The fabricated nanopillars can be programmed into the deformed state, by applying stress at an elevated temperature where structures are in the rubbery state. After cooling down to the glassy state, the external stress can be removed, leaving structures in the deformed configuration. Reheating the structures above the glass transition temperature of the material recovers the nanopillars to their original state due to the shape memory effect[25,26], overcoming the VDW forces between pillars in contact with each other or the substrate in the deformed state. A high storage modulus at the rubbery state is needed to enhance the restoring force and avoid lateral collapse of the pillar. Additionally, to achieve nanoscale resolution by TPL, the materials should polymerize fast upon light source exposure with a writing speed of ~0.1 mm/s or higher to avoid overheating and bubbling caused by the high energy femtosecond laser.

To meet these requirements, we designed a shape memory polymer resin based on a copolymer system of an acrylic acid (AAc) and 2-hydroxy-3-phenoxypropyl acrylate (HPPA) (Fig. 1aII). Poly AAc forms hydrogen bonds and serves as the stiff chain. Poly HPPA functions as an elastomer at room temperature[27], offers the flexibility needed to sustain large deformation in the set state of the pillars. A strong crosslinker dipentaerythritol penta-/hexa-acrylate (DPEPA), containing multi-branched acrylate functional groups (Supplementary Fig. 1) was adopted to impart the rigidity needed to overcome the VDW force during the recovery process and facilitate the polymerization process to pattern the polymer. To eliminate the influence of hydrogen bonds on the surface, a layer of (heptadecafluoro-1,1,2,2-tetrahydrodecyl) trimethoxysilane was coated on the surface after fabrication (Methods) to make it hydrophobic and decrease the surface energy. The low peak values of the loss tangent, tan $\delta$ (ratio between loss modulus and storage modulus) ranging from 0.40-0.61 (Fig. 1b), indicate that the material effectively stores the strain energy, which it can later use to overcome VDW force during the recovery stage. This material system thus exhibits a higher degree of elasticity at the glass transition temperature in contrast to traditional SMPs with peak tan $\delta$ larger than 1. By adjusting the proportion of the chemical composition, a wide range (52 °C to 106 °C) of glass transition temperature ($T_g$) (determining from the peak of tan $\delta$) (Fig. 1b) can be achieved. The storage modulus of the polymer increases with increasing concentration of the acrylic acid (Fig. 1c). High storage moduli of ~2.4-5.2 GPa at 22 °C and ~22.7-90 MPa in the rubbery state were observed, as shown in Fig. 1d. A mass ratio of 1:1:0.8 (AAc, HPPA and DPEPA) was used throughout this manuscript unless otherwise specified, with storage modulus of 5.2 GPa at 22 °C and ~90 MPa at 150 °C in the rubbery state. This storage modulus in the rubbery state is almost an order of magnitude higher than those reported for soft SMPs, and even ~2x higher than fiber reinforced polymers (Fig. 1e). Where necessary, the storage modulus can be further increased by increasing the ratio of crosslinker or decreasing the amount of elastomer. By preparing a photoresist based on this composition (Methods) and patterning by TPL (Methods, Supplementary Fig. 2), nanopillars with aspect ratios of ~8-10 and pitches of 1.2-2.0 μm can be easily fabricated (Fig. 1f). Here the aspect ratio is defined as the height divided by the diameter in the middle part of the nanopillar (Supplementary Fig. 3).

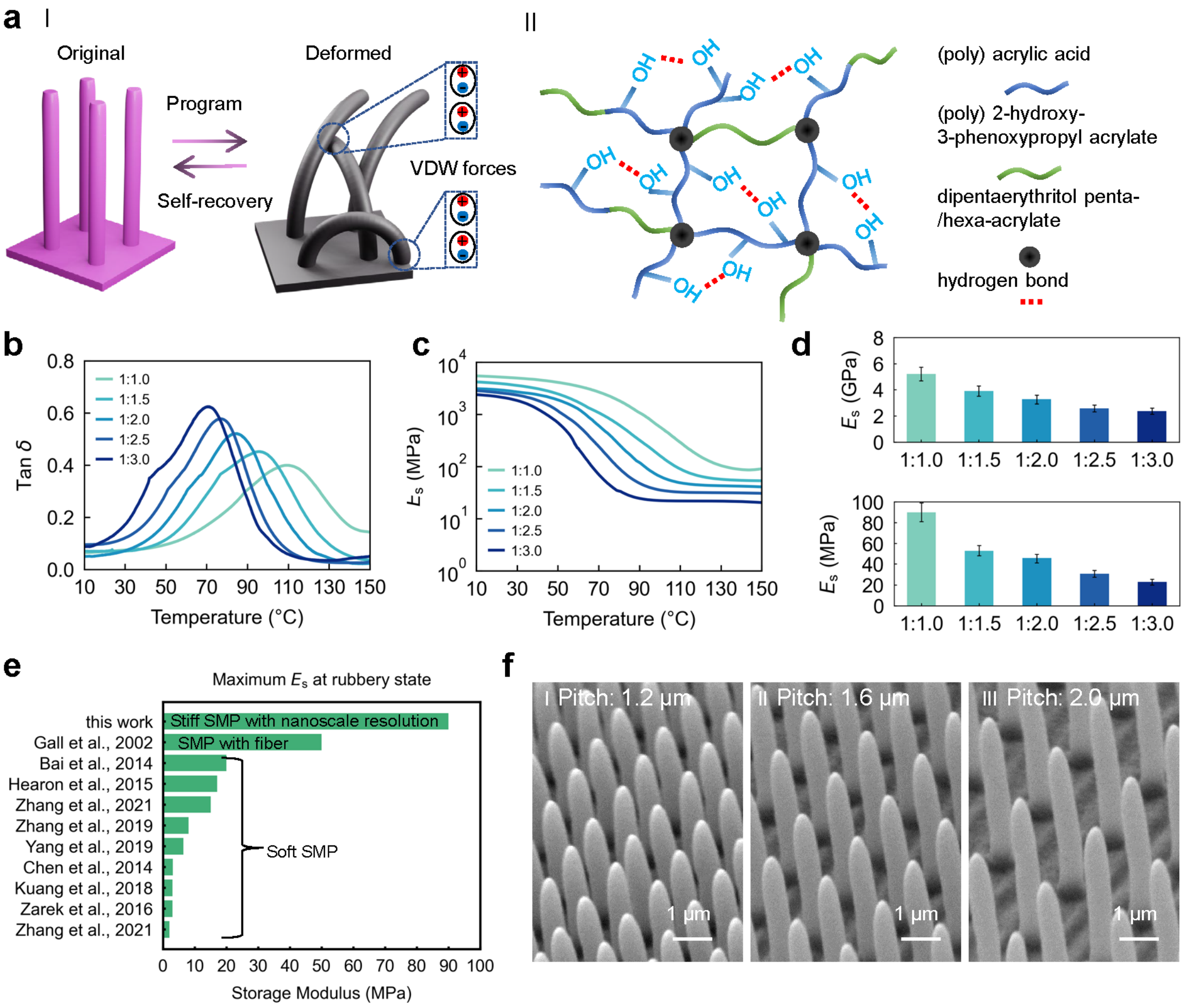

**Fig. 1 Design, characterization, and fabrication of nanostructures in stiff self-recovery materials. a** (I) The concept of reconfigurable transformation of free-standing nanopillars. (II) Design of the Acrylic acid (AAc) and 2-hydroxy-3-phenoxypropyl acrylate (HPPA) copolymer system. **b** Plot of Tan δ (the ratio between loss modulus and storage modulus) as a function of temperature for different compositions (shown as ratio between AAc to HPPA, with DPEPA kept at 0.8). **c** Storage modulus as a function of temperature for different compositions. **d** Storage modulus at room temperature (22 °C, top) and high temperature (40 °C above the glass transition temperature, bottom) for different compositions. Values in **d** represent mean and the error bars represent the standard deviation of the measured values ($n$=3). **e** A comparison of the storage modulus of the SMPs in rubbery

state from different reported works (references in this panel are given in Supplementary Table 1). **f** Tilted view (45°) scanning electron micrographs (SEM) of as fabricated nanopillar arrays with varying pitches. Nanopillars have a nominal diameter of 400 nm and heights of 3.3 µm, 3.9 µm and 3.9 µm respectively (aspect ratio of ~8-10).

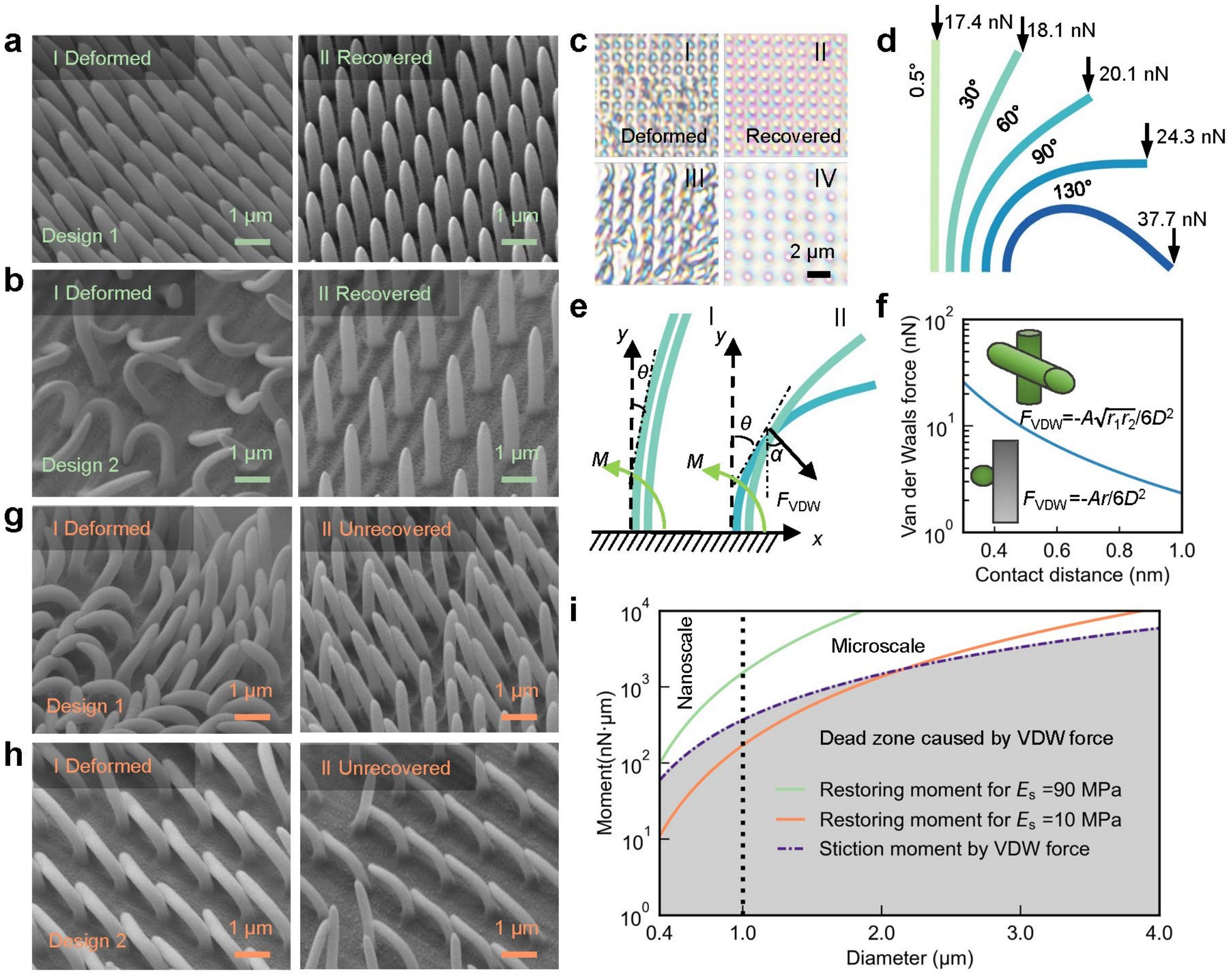

**Fig. 2 Study of recoverability of stiff vs soft nanopillars. a-b** Tilted view (45°) SEM images of the deformed (I) and recovered (II) stiff nanopillars ($E_s$=90 MPa) for in dense and sparse arrays (design 1 and 2). **c** Corresponding optical images of the deformed (I and III) and recovered (II and IV) nanopillars. **d** Calculated force to bend a nanopillar (diameter: 400 nm, height: 4 µm) to different angles and corresponding bent shape. **e** Force analysis of two nanopillars in the bent state for two separated (I) and touching (II) pillars. *M* represents restoring moment of bent pillars, $\theta$ is the bending angle, $\alpha$ is the direction of the VDW force $F_{VDW}$. **f** Calculated VDW force as a function of the contact distance for both crossed cylinders (upper inset) and sphere-wall (lower inset) cases. These two cases are equivalent if the two pillars and the pillar tip hemisphere have the same diameter. *A* is the Hamaker constant, $r_1$ and $r_2$ are the radius of the two pillars, *r* is the radius of the pillar tip, *D* is the contact distance. **g-h** Tilted view (45°) SEM images of the deformed (I) and unrecovered (II) soft nanopillars ($E_s$=10 MPa) for design 1 and 2 respectively. **i** Calculated restoring moment and stiction moment by VDW force as a function of pillar diameter for a fixed aspect ratio of 10 and bending angle of 90° with VDW force acting on the tip of the pillar showing dominance of VDW forces for structures with diameters smaller than ~2.2 µm for nanostructures in materials of low storage modulus of 10 MPa. Increasing storage modulus to 90 MPa enables sub-micron reconfigurable nanostructures as restoring moment exceeds VDW moment. The grey region represents the 'dead zone' where pillars with restoring moment in this region cannot recover.

To study the self-recovery of the stiff material, we 3D printed nanopillars with aspect ratio of ~10 in dense and sparse arrays, denoted as design 1 and design 2 with pitches of 1.2 and 2 μm respectively. The nanopillars were programmed by compressing at 126 °C (20 °C above $T_g$) with a pressure of 100 psi using a Nanonex nanoimprint machine (Methods, Supplementary Fig. 4). In both designs, the pillars were bent in random directions resulting in some pillars touching each other (Fig. 2 aI and Fig. 2 bI). Cooling the nanostructures to room temperature with constant pressure applied stores it in the deformed state as the polymer chains remain kinetically trapped even after the stress is lifted. Heating the sample to 170 °C brings the pillars into the rubbery state again, triggering the distorted pillars to recover to their original standing state (Fig. 2 aII and Fig. 2 bII). In this rubbery state, the stored energy is released and its restoring force exceeded the VDW force between the touching pillars. The structures of the same sample can be directly observed under an optical microscope before and after recovery (Fig. 2c). The shape recovery process completed within 60 s once heated above the glass transition temperature (Supplementary Video 1). Using classical elasticity theory, we calculated the forces acting at the tip of a single pillar to bend it to different angles, and its corresponding bent shape (Fig. 2d, Supplementary part 6). These shapes agree with those observed in the SEMs of deformed nanopillars. During the self-recovery process, isolated pillars can recover freely due to the restoring moment $M$ at the base and along the length of the pillar (Fig. 2eI, Supplementary Equation S4). When the pillars are touching, the restoring moment $M$ needs to be larger than the moment caused by the VDW force between pillars (Fig. 2eII). Thus, a higher storage modulus produces higher restoring moments leading to nanostructure recovery.

We considered VDW forces in two commonly observed cases: between two nanopillars, and between the tip of a nanopillar and the base of the print. Assuming that nanopillars have equal diameters, the VDW stiction force for both cases are the same as described in Supplementary Equation S5 and plotted versus contact distance as shown in Fig 2f. We then consider how the restraining stiction moment provided by the VDW force competes against the restoring moment of a soft pillar (storage modulus of 10 MPa) and a stiff pillar (storage modulus of 90 MPa). The respective restraining and restoring moments were calculated for a 90° bending angle and contact distance of ~0.3 nm[15] and plotted as a function of pillar diameter in Fig 2i (see Supplementary part 6 for calculations). At diameters smaller than ~2.2 μm, the restoring moment for the soft pillar falls into the 'dead zone' (grey region) where it is lower than the stiction moment. In contrast, the restoring moment of the stiff pillar is higher than the stiction moment for all pillar diameters down to 400 nm, which corresponds to the nominal diameter of nanopillars in this study.

A control experiment was conducted to examine the self-recovery effect of pillars made from the soft material (Supplementary part 7). With the same design and fabrication process as in Fig. 2a-b for the stiff pillars, the soft pillars that touch could no longer recover due to the stickiness caused by the VDW force (Fig. 2g-h), indicating the advantage of the stiff material. Note that for designs of other reversible geometries at nanoscale, this effect of stiction will need to be considered. To examine this, we printed tall gratings instead of pillars (Supplementary Fig. 6) and observed a similar trend, i.e., the structures of stiffer material can overcome the VDW stiction force where the softer counterpart remain permanently stuck. The shape recovery might depend on the pillar aspect ratio and pitch as seen in larger structures[19]. To investigate these two factors, pillars with different aspect ratios and pitches were fabricated and programmed. The degree of recovery remains at 100% for aspect ratio no larger than 8 for all pitches (Supplementary Fig. 7). Here, the recovery ratio was defined as $n_{\text{recovered}}/n_{\text{total}}$, where $n_{\text{recovered}}$ is the number of pillars recovered to the straight state, and $n_{\text{total}}$ is the total number pillars per printing area (20μm×20μm). Higher aspect ratio leads to partial recovery, especially along the edges of the printed area (Insets of Supplementary Fig. 7), which may be caused by the stress concentration along the edges during the compression procedure. For pitch of 1.2 μm, aspect ratio larger than 10 was not examined because the pillars would collapse due to capillary forces during the development and drying process (Supplementary Fig. 8).

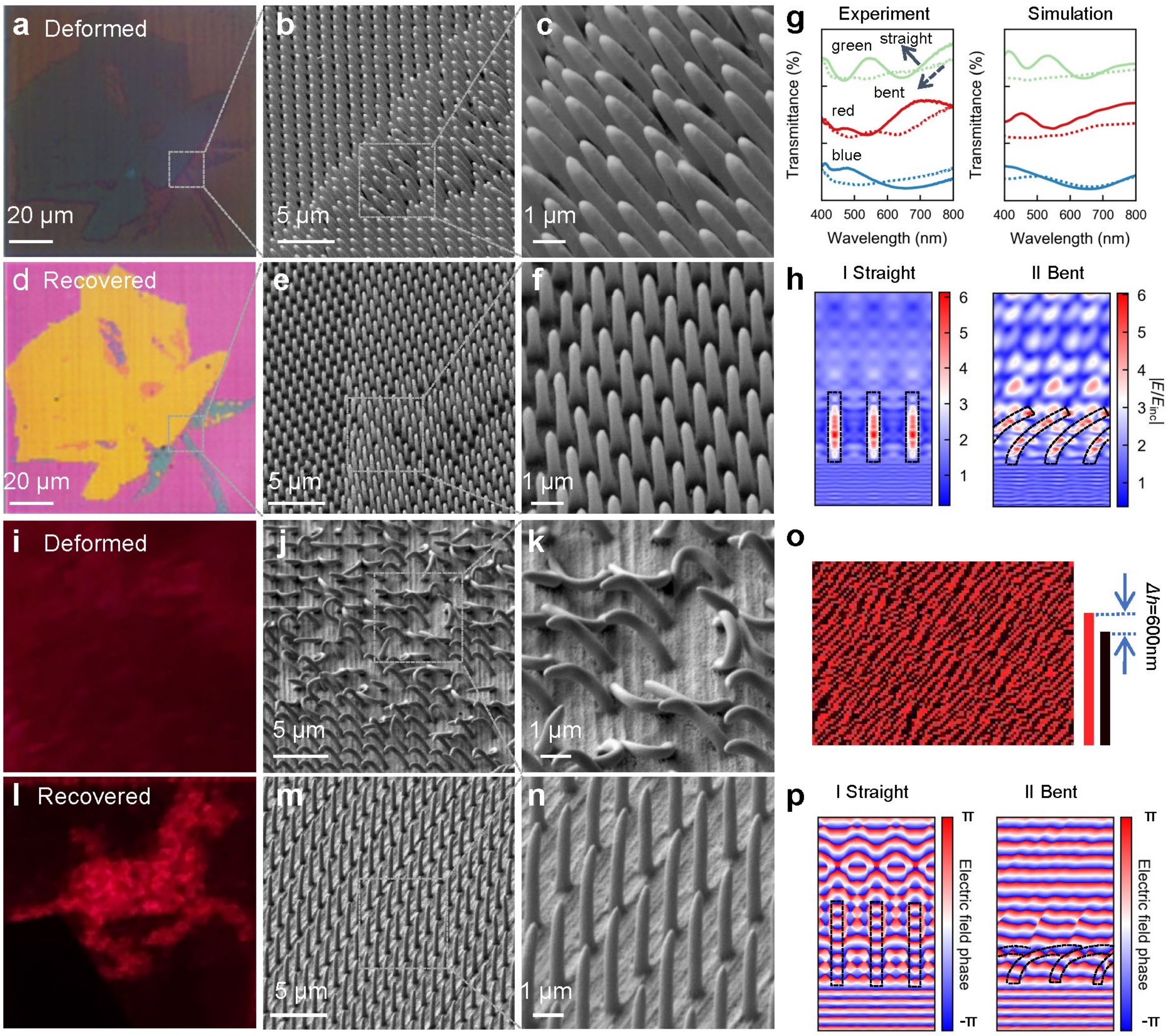

**Fig. 3 Applications of the stiff self-recovery nanopillars in nano optics. a** Optical transmittance micrograph of a deformed color microprint consisting of nanopillars observed under an objective lens (NA=0.2, CA=11.5°). **b-c** Tilted view (45°) SEM images of the programmed painting. **d-f** Optical transmittance micrograph and corresponding tilted view (45°) SEM images of the recovered painting. **g** Measured and simulated transmittance spectra of blue, red, and green nanopillars at the straight (solid lines) and bent (dashed lines) states. **h** Cross section views of the nearfield normalized electric field amplitude at 550 nm wavelength for green nanopillars (at the straight and bent states respectively). **i** A far field holographic projection from a sample of deformed nanopillars under a 635 nm red laser illumination showing noise **j-k** Tilted view (45°) SEM images of the deformed painting. **l-n** Projection and corresponding tilted view (45°) SEM images of a recovered hologram respectively. **o** Binary phase map of the designed hologram. **p** Cross sectional views of the nearfield phase maps at 635 nm wavelength for nanopillars in the straight and bent states.

Due to the interaction of the nanopillars with visible light, these programmable nanopillars could exhibit interesting optical effects. Different structural colours[21] can be printed by changing the number of printing layers and exposure time (Supplementary Fig. 9). An image of a flower was printed with a pixel size of 1.2 μm (Fig. 3d) and a resolution of ~21,150 dots per inch (d.p.i.). This pixel resolution is ~8.3 times higher than our previous study on nanoscale patterning of SMPs[3], benefiting from the single nanopillar pixel adopted. As a single nanopillar[22] can generate vivid structural colours, the crosstalk between adjacent pixels is strongly suppressed. By the programming process, the as printed structures can be programmed into a dark patch (Fig. 3a), with the pillars bending and some parts touching each other (Fig. 3b-c). Upon heating to the elevated temperature (170 °C), the print recovered to its original state to reveal an image (Fig. 3d), with the bent and connected pillars returning to the straight as printed state (Fig. 3e-f).

We simulated the bending of the pillar using Finite Element Analysis (FEA) (Methods, Supplementary Fig. 10) and used the resulting shapes in finite difference time domain (FDTD) simulation to compute the transmittance spectra of the pillar under white light illumination. Both experimental measurement and FDTD simulation show that the spectra of blue, red, and green colours were "flattened" in the bent state (Fig. 3g), leading to the grey appearance in Fig. 3a. Near field electric field amplitude at wavelength of 550 nm shows a wave guiding mode within the straight nanopillar with light illuminated vertically from the bottom (Fig. 3hI), while complex interference patterns is observed within the nanopillar and air in the bent state (Fig. 3hII), resulting in weaker wavelength-selective effect, hence flattening the peaks in the spectra (Fig. 3g). More structural colours at the deformed and recovered states are given in Supplementary Fig. 11a. A comparison of the spectra of different structural colours recovered from multiple programming cycles shows nearly full recovery compared to the original spectra (Supplementary Fig. 11b). Note that for simplicity, a parallel configuration was adopted for simulating the spectra in the bent state. A comparison of the parallel and touched configurations (Supplementary Fig. 11c) shows that they have similar spectra devoid of well-defined peaks and throughs. The tuning of light amplitude can be extended to near-infrared (NIR) light range (Supplementary Fig. 12), using the continuous mode in TPL to achieve larger diameters of the nanopillars (Supplementary Fig. 13).

Apart from controlling the amplitude of light, the nanopillars also manipulate the phase of light. A binary hologram of a horse was designed using pillars with height difference of ~600 nm to generate ~$\pi$ phase difference at 635 nm wavelength (Fig. 3o, Supplementary Fig. 14). In the deformed state, the hologram cannot be seen (Fig. 3i), indicating complete deformation of the entire area of the hologram (Fig. 3j-k). Heating led to the recovery of the hologram (Fig. 3l) with the pillars in the recovered state (Fig. 3m-n). The simulated phase of a pillar indicates an erasure of the phase information caused by the nanopillar in the bent state (Fig. 3p), leading to the disappearance of the hologram.

Van der Waals (VDW) force are easily overcome in reversible materials and structures at the macroscale, where the internal forces are sufficiently large to detach surfaces that come in contact. However, we observe from our theoretical analysis and experiments that nanostructures of the storage modulus of <20 MPa in typical SMPs result in structures that are too soft to overcome VDW forces at sub-micron dimensions. Stiffness of sub-micron structures can be increased by incorporating a larger amount of crosslinkers, resulting in SMPs with unprecedentedly high storage modulus approaching 100 MPa even while in the rubbery state. This added stiffness imparts high-aspect ratio structures the ability to detach from each other in the dense array (design 1) or to detach from the substrate in the sparse array (design 2). The structures remain highly deformable as can be seen in the curled up pillars in the set state without fracturing.

The overall dark appearance of bent nanopillars in Fig. 3a is in stark contrast to the transparent appearance of the compressed mesh structures consisting of softer SMPs in the overly constrained configuration[3]. As the pillars are isolated structures that act also as weak waveguides, light scatters at the interface with the surrounding trapped air between bent pillars, causing a decreased transmittance. Similarly, the disappearance of any recognizable pattern in the hologram is a result of the severely and randomly curled up pillars.

The formulated material system not only produces high aspect ratio structures that can recover from large deformation and collapse, but also functions as a high resolution resin enabling sub-micron 4D printing[7] of other complex geometries. Sub-micron structures that would otherwise remain permanently deformed can now overcome VDW forces and exhibit reconfigurability. With sub-micron features, interactions with light in the visible spectrum becomes possible as demonstrated by the structural colours produced with ~21,150 dpi resolution. By controlling the amplitude and phase of light in the upright and collapsed state, we achieve erasable-recoverable colour prints and diffractive optical elements potentially useful as microscopic temperature recording labels. Future works may consider the applications in NIR range including tunable infrared imaging[28,29] and radiative cooling[30], while large scale fabrication technologies such as nanoimprint[31] and self-assembly[32] should be implemented. We believe that our approach here could benefit the further miniaturization of reconfigurable mechanical devices in various research fields.

## Methods

**Materials.** Acrylic acid (AAc) (anhydrous, contains 200 ppm MEHQ as inhibitor, 99%), 2-hydroxy-3-phenoxypropyl acrylate (HPPA) (contains 250 ppm monomethyl ether hydroquinone as inhibitor), Dipentaerythritol penta-/hexa-acrylate (DPEPA) (contains ≤650 ppm MEHQ as inhibitor), Polyvinylpyrrolidone (PVP) (average molecular wight~1300000), Acetone (≥99.5%), isopropyl alcohol (IPA) (99.7%), ethanol (anhydrous, ≥99.5%), 3-(Trimethoxysilyl) propyl methacrylate (98%), Acetic acid (≥ 99%), diphenyl (2,4,6-trimethylbenzoyl) phosphine oxide (TPO) were purchased from Sigma Aldrich. 7-diethylamino-3-thenoylcoumarin (DETC) was purchased from Exciton. (Heptadecafluoro-1,1,2,2-tetrahydrodecyl) trimethoxysilane was purchased from Gulf Chemical. All the chemicals were used as received.

**Preparation of photoresist.** 1 g of HPPA and 0.8 g of DPEPA were added in 1 g of AAc and stirred by a magnetic rotor for 10 mins to get homogeneous solution. Then 30 mg of DETC was added into the solution and stirred for another 30 mins. Finally, 150 mg of PVP was added into the solution and stirred on a hotplate (40 °C) overnight (8 hours) to get the photoresist.

**Structure fabrication.** Before printing, a borosilicate glass ($\Phi 25\times 0.17$ $mm^3$) was cleaned with IPA solution in an ultrasonic cleaner for 3 mins and dried with nitrogen. To increase adhesion, the cleaned glass was immersed in a solution consists of 47.5 g ethanol, 2.5 g of deionized water, 1.5 g of 3-(Trimethoxysilyl) propyl methacrylate and 15 mg of acetic acid for 10 mins. Then the glass was washed by IPA and dried with nitrogen again. The glass was fixed on the sample holder by adhesive tape and a drop of objective oil (refractive index =1.518, Nikon) was placed onto one side of the glass while the photoresist was on the other side. Then the fixed glass was transferred to a two-photon lithography system (Photonic Professional GT, Nanoscribe GmbH, Germany). A 63×NA1.4 objective lens in the oil immersion configuration (Supplementary Fig. 2) was used. For printing of the substrate, a fixed laser power (15 mW) and write speed (10 mm/s) were used in the continuous writing mode. For printing of the nanopillar with diameter of 400 nm, the laser power and exposure time were set as 15 mW and 0.6-1.2 ms respectively in the pulse writing mode. Nanopillars with diameter larger than 400 nm were printed using the continuous mode and the same printing parameters as the base. To rinse the uncured photoresist after printing, the glass was immersed in acetone and IPA for 5 mins sequentially, then in acetone, IPA, and deionized water in the ultrasonic cleaner for 1 min respectively. Afterwards, the sample was taken out from the deionized water and dried with nitrogen. After drying, the glass was treated in oxygen plasma for 10 seconds, and coated with (heptadecafluoro-1,1,2,2-tetrahydrodecyl) trimethoxysilane at 80 °C for 1 hour in an integrated Cleaner/Coater (Ultra-100 Nanonex).

**Materials characterisation.** The dynamic mechanical analysis tests to measure the storage modulus and tan $\delta$ at different temperature were conducted on a DMA machine (TA Instruments, Q800, U.S.) in the tension film mode, started from 150 to 10 °C at a cooling rate of 2 °C/min. The samples (6 mm × 15 mm × 0.5 mm) were made by curing the photoresist in a Teflon mould in a UV oven for 10 mins. In this part, to cure the photoresist by the UV light source, the initiator diphenyl (2,4,6-trimethylbenzoyl) phosphine oxide (TPO) instead of DETC was used at a mass concentration of 2%. Meanwhile, PVP which does not participate in the chemical polymerization process was not added to avoid too high viscosity when moulding the samples.

The programming process was conducted under a Nanonex NX-2006 nanoimprint machine. The sample was put between two pieces of big Teflon films held by a holder. Two pieces of silicon wafer was put between these two big Teflon films to make the sample receive the pressure more uniformly (Supplementary Fig. 4). A piece of small Teflon film was placed above the sample to avoid contamination of the nano structure during the programming process. Every process involved pumping down at room temperature (22 ℃) to remove the air trap and ensure sample-substrate conformity, followed by heating the films to a high temperature (20 ℃ above the material's glass transition temperature). At the high temperature, pressure (100 psi for the stiff material and 10 psi for the soft control material) was applied on the films for 1 min. With the pressure on, the films were cooled down to room temperature. At room temperature, the pressure was released, and the sample was taken out for further characterization. The shape recovery process was activated by heating the sample to above the glass transition temperature either by a heating gun or heating stage (Linkam Scientific Instruments Ltd).

Transmittance spectra at the visible light range (400-800 nm) were measured using an objective lens (NA=0.2, CA=11.5°) on an optical microscope (Nikon Eclipse LV100ND) with a CRAIC 508 PV microspectrophotometer and a Nikon DS-Ri2 camera.

Transmittance spectra at NIR range were measured with an inverted optical microscope (Nikon Ti-2) coupled to a spectrometer (Andor Kymera-328i) equipped with an InGaAs detector (Andor iDus- DU490A). The samples were excited by a top halogen lamp collimated by a condenser with a minimally closed iris at its back focal plane, giving a numerical aperture of ~ 0.02. The signals were then collected by a bottom objective (Nikkon 50x, 0.55NA) and dispersed onto 150 lines/mm grating resulting in a spectral resolution of ~ 1.5 nm.

The uniaxial tensile experiments to get the constitutive model for FEA simulation were conducted on the DMA tester (TA Instruments, Q800, U.S.) in the stress control mode with a stress rate of 2 MPa/min at 126 ℃.

The refractive index of the photoresist (Supplementary Fig. 15) used in the FDTD simulation was measured by an EP4 Ellipsometer (ACCURION, Germany). To prepare the sample for measurement, the photoresist (without adding PVP to avoid too high viscosity) were diluted by IPA with a mass ratio of 1:9 and spun coated on a silicon wafer at 5000 rpm for 10 seconds. The IPA was evaporated during and after the coating process. Afterwards, the sample was cured by UV light for 3 mins.

**Numerical Simulation.** FDTD simulation was conducted with a commercial software (FDTD, Lumerical Solutions). The dimension profile for the simulation was obtained from the SEM (JEOL, Japan) images and measured using the ImageJ software. The deformed configuration of the nanopillar used in the FDTD study was calculated using the structural mechanics module in COMSOL Multiphysics. The constitutive equation of the material was obtained by fitting the tensile test data with the Mooney-Rivlin hyperelastic model.

## References

1 Li, S. *et al.* Liquid-induced topological transformations of cellular microstructures. *Nature* **592**, 386-391 (2021).

2 Xia, X. *et al.* Electrochemically reconfigurable architected materials. *Nature* **573**, 205-213 (2019).

3 Zhang, W. *et al.* Structural multi-colour invisible inks with submicron 4D printing of shape memory polymers. *Nat. Commun.* **12**, 1-8 (2021).

4 Lu, Y., Aimetti, A. A., Langer, R. & Gu, Z. Bioresponsive materials. *Nat. Rev. Mater.* **2**, 1-17 (2016).

5 Truby, R. L. & Lewis, J. A. Printing soft matter in three dimensions. *Nature* **540**, 371-378 (2016).

6 Zarek, M. *et al.* 3D printing of shape memory polymers for flexible electronic devices. *Adv. Mater.* **28**, 4449-4454 (2016).

7 Ge, Q. *et al.* Multimaterial 4D printing with tailorable shape memory polymers. *Sci. Rep.* **6**, 1-11 (2016).
8 Jin, D. *et al.* Four-dimensional direct laser writing of reconfigurable compound micromachines. *Mater. Today* **32**, 19-25 (2020).
9 Hippler, M. *et al.* Controlling the shape of 3D microstructures by temperature and light. *Nat. Commun.* **10**, 1-8 (2019).
10 Ford, M. J. *et al.* A multifunctional shape-morphing elastomer with liquid metal inclusions. *Proc. Natl. Acad. Sci. U.S.A.* **116**, 21438-21444 (2019).
11 Ze, Q. *et al.* Magnetic shape memory polymers with integrated multifunctional shape manipulation. *Adv. Mater.* **32**, 1906657 (2020).
12 Koenderink, A. F., Alu, A. & Polman, A. Nanophotonics: Shrinking light-based technology. *Science* **348**, 516-521 (2015).
13 Guo, L. *et al.* Strategies for enhancing the sensitivity of plasmonic nanosensors. *Nano Today* **10**, 213-239 (2015).
14 Zeng, H., Wasylczyk, P., Wiersma, D. S. & Priimagi, A. Light robots: bridging the gap between microrobotics and photomechanics in soft materials. *Adv. Mater.* **30**, 1703554 (2018).
15 Autumn, K. *et al.* Adhesive force of a single gecko foot-hair. *Nature* **405**, 681-685 (2000).
16 Duan, H. & Berggren, K. K. Directed self-assembly at the 10 nm scale by using capillary force-induced nanocohesion. *Nano Lett.* **10**, 3710-3716 (2010).
17 Li, M., Tang, H. X. & Roukes, M. L. Ultra-sensitive NEMS-based cantilevers for sensing, scanned probe and very high-frequency applications. *Nat. Nanotechnol.* **2**, 114-120 (2007).
18 Perutz, S., Wang, J., Kramer, E., Ober, C. & Ellis, K. Synthesis and surface energy measurement of semi-fluorinated, low-energy surfaces. *Macromolecules* **31**, 4272-4276 (1998).
19 Chen, C. M. & Yang, S. Directed water shedding on high‐aspect‐ratio shape memory polymer micropillar arrays. *Adv. Mater.* **26**, 1283-1288 (2014).
20 Elliott, L. V., Salzman, E. E. & Greer, J. R. Stimuli Responsive Shape Memory Microarchitectures. *Adv. Funct. Mater.* **31**, 2008380 (2021).
21 Wang, H. *et al.* Full color and grayscale painting with 3D printed low-index nanopillars. *Nano Lett.* (2021).
22 Chan, J. Y. E. *et al.* High-resolution light field prints by nanoscale 3D printing. *Nat. Commun.* **12**, 1-9 (2021).
23 Nawrot, M., Zinkiewicz, Ł., Włodarczyk, B. & Wasylczyk, P. Transmission phase gratings fabricated with direct laser writing as color filters in the visible. *Opt. Express* **21**, 31919-31924 (2013).
24 Wang, H. *et al.* Optical Fireworks Based on Multifocal Three-Dimensional Color Prints. *ACS Nano* (2021).
25 Xie, T. Tunable polymer multi-shape memory effect. *Nature* **464**, 267-270 (2010).
26 Jin, B. *et al.* Programming a crystalline shape memory polymer network with thermo- and photo-reversible bonds toward a single-component soft robot. *Sci. Adv.* **4**, eaao3865 (2018).
27 Zhang, B., Kowsari, K., Serjouei, A., Dunn, M. L. & Ge, Q. Reprocessable thermosets for sustainable three-dimensional printing. *Nat. Commun.* **9**, 1-7 (2018).
28 Ran, W. *et al.* An integrated flexible all‐nanowire infrared sensing system with record photosensitivity. *Adv. Mater.* **32**, 1908419 (2020).

29 Tang, X., Ackerman, M. M., Chen, M. & Guyot-Sionnest, P. Dual-band infrared imaging using stacked colloidal quantum dot photodiodes. *Nat. Photonics* **13**, 277-282 (2019).
30 Shi, N. N. *et al.* Keeping cool: Enhanced optical reflection and radiative heat dissipation in Saharan silver ants. *Science* **349**, 298-301 (2015).
31 Cox, L. M. *et al.* Nanoimprint lithography: Emergent materials and methods of actuation. *Nano Today* **31**, 100838 (2020).
32 Cummins, C. *et al.* Enabling future nanomanufacturing through block copolymer self-assembly: A review. *Nano Today* **35**, 100936 (2020).

**Acknowledgements**

This research was supported by National Research Foundation (NRF) Singapore, under its Competitive Research Programme award NRF-CRP20-2017-0004 and NRF Investigatorship Award NRF-NRFI06-2020-0005.

**Author Contributions**

W.Z. conceived the idea, designed the experiments, developed the photoresists, fabricated, and characterized the samples with the assistance from H.W. and J.K.W.Y. H. W. performed the FDTD simulation. A. T. L. T. and W. Z. did the elasticity theory calculation. W. Z. did the FEA simulation. A. S. R., B. Z., H. T. W., J. Y. E. C., Q. R., H. L., S. T. H., L. D., D. W., V. K. R., H. Y. L. assisted the characterization. J.K.W.Y. supervised the research. All authors contributed to writing and revision of the manuscript.

**Competing interests**

The authors declare no competing interests.

**Additional information**

Supplementary Information is available for this paper.

**Correspondence** and requests for materials should be addressed to H.W. or J.K.W.Y.

Supplementary information for

# Overcoming Van der Waals Forces in reconfigurable nanostructures

Wang Zhang[1], Hao Wang[1*], Alvin T. L. Tan[1], Anupama Sargur Ranganath[1], Biao Zhang[2], Hongtao Wang[1], John You En Chan[1], Qifeng Ruan[1], Hailong Liu[3], Son Tung Ha[3], Dong Wang[4], Venkat K. Ravikumar[5], Hong Yee Low[1], Joel K.W. Yang[1,3*]

[1]Engineering Product Development, Singapore University of Technology and Design, Singapore 487372, Singapore. [2]Frontiers Science Center for Flexible Electronics, Xi'an Institute of Flexible Electronics (IFE) and Xi'an Institute of Biomedical Materials & Engineering (IBME), Northwestern Polytechnical University, 127 West Youyi Road, Xi'an 710072, China. [3]Institute of Materials Research and Engineering, Singapore 138634, Singapore. [4]Robotics Institute and State Key Laboratory of Mechanical System and Vibration, School of Mechanical Engineering, Shanghai Jiao Tong University, Shanghai 200240, PR China. [5]Advanced Micro Devices Singapore Pte Ltd, Singapore.

*Email: whchn@live.cn; joel_yang@sutd.edu.sg

## 1. Preparation of the photoresist

1.5 g of 2-hydroxy-3-phenoxypropyl acrylate (HPPA) and 0.8 g of dipentaerythritol penta-/hexa-acrylate (DPEPA) were added in 1 g of acrylic acid (AAc) and stirred by a magnetic rotor for 10 mins to get homogeneous solution. Then 30 mg of DETC was added into the solution and stirred for another 10 mins. Finally, 150 mg of PVP was added into the solution and stirred on a hotplate (40 °C) overnight to get the photoresist.

In this photoresist, AAc serves as the stiff chain, and HPPA was used to form copolymer with AAc to offer the flexibility of the molecular chain needed during the programming process. DPEPA, containing multi-branched acrylate functional groups, was adopted to increase the mechanical stability and facilitate the two photon polymerisation process to pattern the shape memory polymer. DETC, a high efficiency photo initiator for two photon lithography[1], was used to ensure the nanoscale resolution needed in this study. PVP (average molecular wight ~1,300,000) was added to adjust the viscosity of the final photoresist.

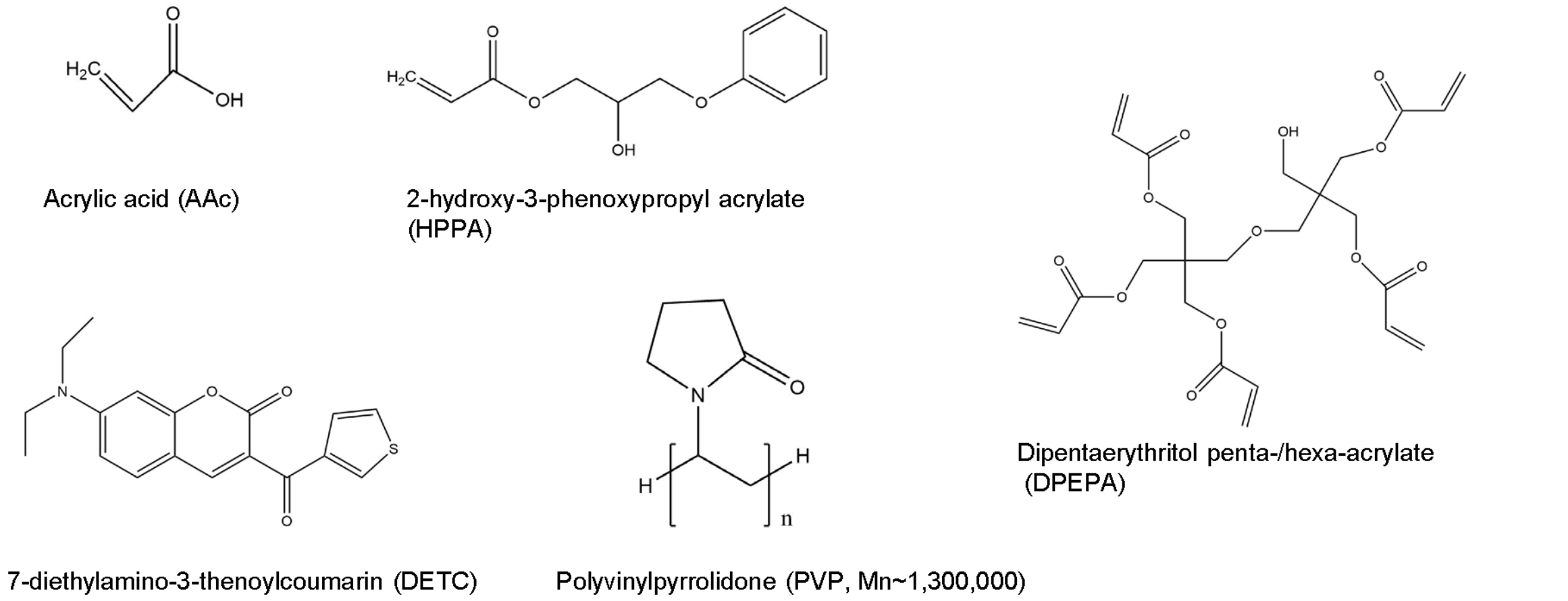

**Supplementary Fig. 1 Chemical structures of the compositions in the photoresist.**

## 2. Two photon lithography

A two-photon lithography system (Photonic Professional GT, Nanoscribe GmbH, Germany) was adopted to fabricate the programmable nanopillars. A 63×NA1.4 objective lens was immersed into the objective oil. The femtosecond laser from the lens passed through the objective oil and the borosilicate glass to cure the photoresist on top of the glass. For printing of the substrate (5-15 printing layers, 200 nm pitching, 200 nm hatching), the continuous laser mode with a laser power 15 mW and a write speed of 10 mm/s was adopted. For printing of the nanopillars, the pulsed laser mode with a laser power of 15 mW and exposure time 0.6-1.2 ms was adopted (3-20 printing layers, 200 nm hatching).

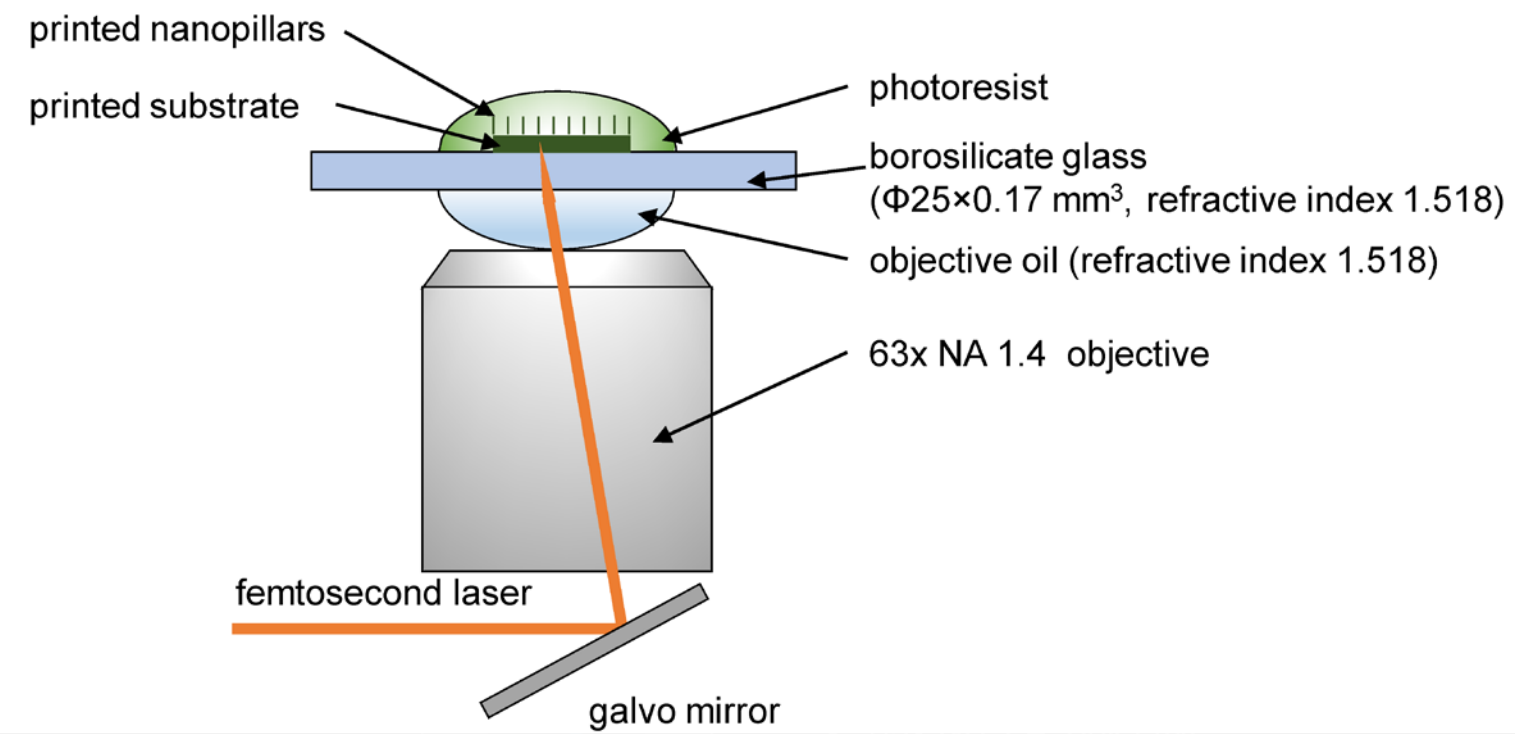

**Supplementary Fig. 2 Schematic of the two photon lithography set up.**

## 3. Determination of the aspect ratio of the nanopillar

The aspect ratio (*AR*) of the nanopillar is defined as the height *h* divided by the diameter *d* in the middle part of the nanopillar. $AR=h/d$. *d* was measured using the imageJ software using the SEM image. *h* can be calculated by $h=h_{measure}/\sin\theta$, where is $h_{measure}$ is the measured height using ImageJ, $\theta$ is the tilted angle of the SEM stage.

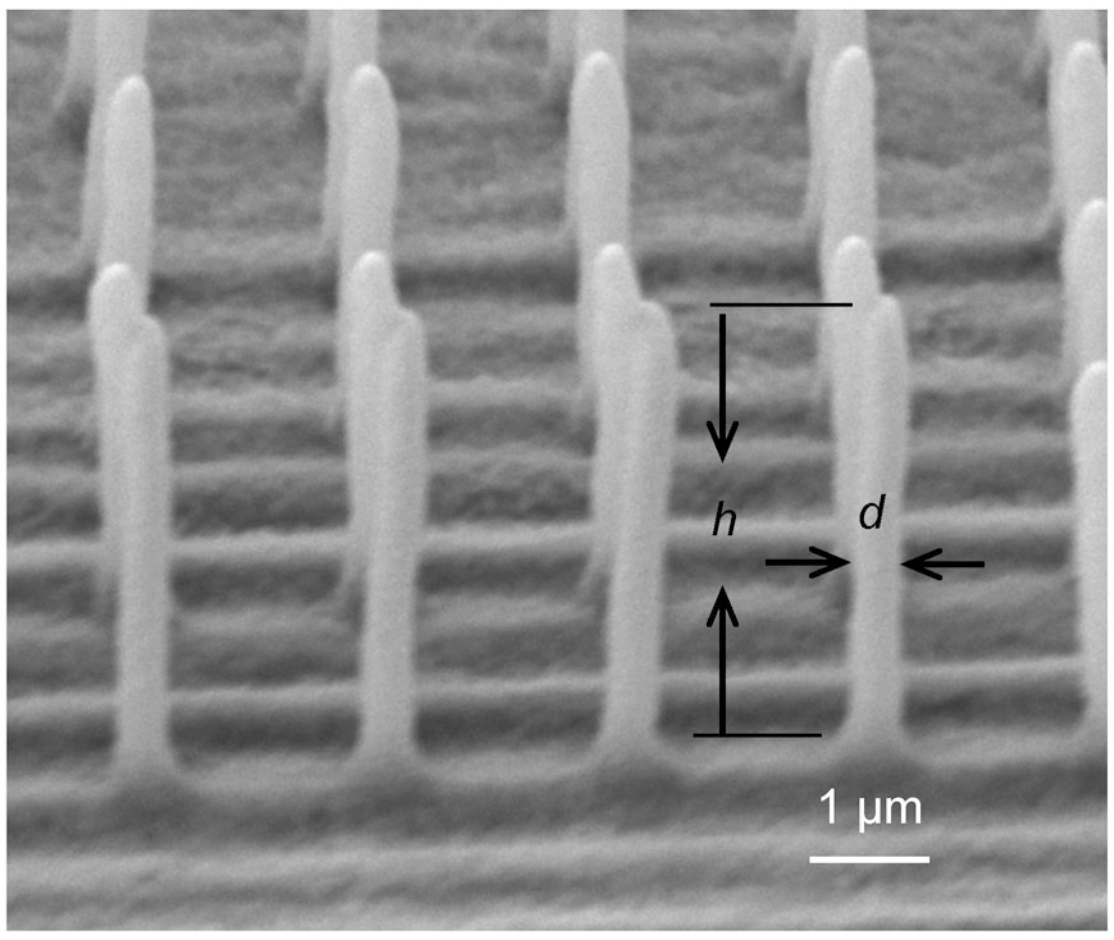

**Supplementary Fig. 3 Tilted view (45°) SEM image as printed nanopillars.**

## 4. Comparison of the storage modulus ($E_s$) at the rubbery state of SMPs

**Table S1. A comparison of the storage modulus and corresponding fabrication processes of the shape memory polymers (SMPs) at rubbery state from different reported works**

| | Maximum $E_s$ at rubbery state (MPa) | Fabrication Process |
|---|---|---|
| **This work** | ~90 | Two-photo lithography |
| **Gall et al.** [2] | ~50 | Moulding |
| **Bai et al.**[3] | ~20 | Moulding |
| **Hearon et al.**[4] | 17.1 | Moulding |
| **Zhang et al.**[5] | 15 | Two-photo lithography |
| **Zhang et al.**[6] | ~8 | Digital light processing |
| **Yang et al.**[7] | 6.4 | Digital light processing |
| **Chen et al.**[8] | 3.1 | Moulding |
| **Kuang et al.**[9] | ~3 | Direct ink writing |
| **Zarek et al.**[10] | ~3 | Digital light processing |
| **Zhang et al.**[11] | ~2 | Digital light processing |

## 5. The programming process

The programming process was conducted under a Nanonex NX-2006 nanoimprint machine. The sample was put between two pieces of big Teflon films held by a holder. Two pieces of silicon wafer were put between these two big Teflon films to make the sample receive the pressure more uniformly. A piece of small Teflon film was placed above the sample to avoid contamination of the nano structure during the programming process. Then the space between the two big films was pumped into vacuum at room temperature (22 °C), followed by heating the films to a high temperature (20 °C above the material's glass transition temperature). At the high temperature, different air pressure was applied on the films for 1 min. With the pressure on, the films were cooled down to room temperature. At room temperature, the pressure was released, and the sample was taken out for further characterization. The shape recovery process was activated by heating the sample to above the glass transition temperature either by a heating gun or heating stage.

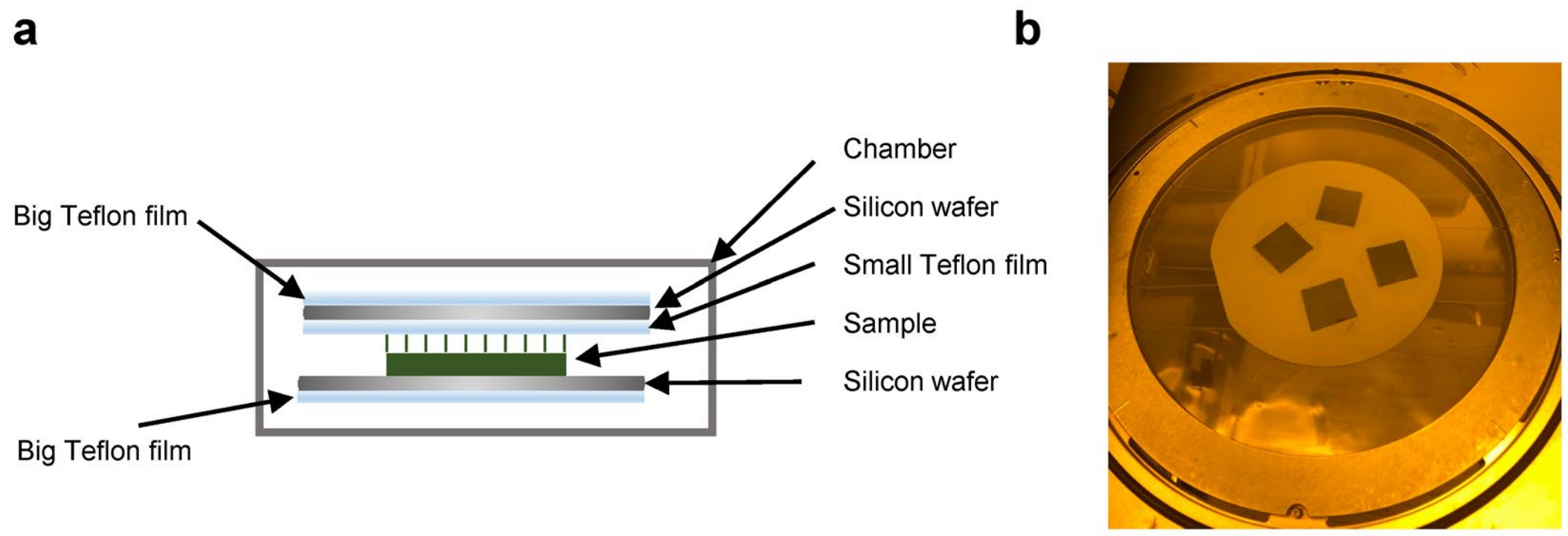

**Supplementary Fig. 4 a** Schematic of the setup of the Nanonex NX-2006 machine. **b** Photo of the setup (with 4 pieces of samples).

## 6. Classical elasticity analysis

The classical elasticity theory was used to get the analytic solution of the bent shape of the nanopillars. Consider a nanopillar with a height of $h$ and a diameter of $d$. At the rubbery state, the force needed to bend the nanopillar to a degree of $\theta_0$ can be calculated as[12]

$$P=\frac{0.5EI}{h^2}\left(\int_0^{\theta_0}\frac{d\theta}{\sqrt{cos\theta-cos\theta_0}}\right)^2 \tag{S1}$$

The shape of the deformed pillar can be described as

$$x=\sqrt{(2EI/P)}[\sqrt{(1-\cos\theta_0)}-\sqrt{(\cos\theta-\cos\theta_0)}] \tag{S2}$$

$$y=\sqrt{(EI/2P)}\int_0^{\theta}\frac{\cos\theta d\theta}{\sqrt{\cos\theta-\cos\theta_0}} \tag{S3}$$

where $E$ is Young's modulus and $E\approx E_s\approx 90$ MPa according to the dynamic mechanical analysis data in Fig. 1c, and $I=\pi d^4/64$ is the moment of inertia.

At the bent state, with the load maintained and cooling the pillar to room temperature, the bending shape can be kept with the load removed because the pillar is at glassy state at room temperature. Upon heating to above the glass transition temperature again, the pillar can recover to the straight state freely if it did not contact with neighbouring pillars, because of the restoring moment

$$M=P\times x \tag{S4}$$

at the base, where $x$ is the deformation in $x$ direction given in Equation S2.

When two pillars are contacted at the bent state, the pillars need to overcome the Van der Waals force $F_{VDW}$ between each other to recover. Considering both crossed cylinders (corresponding to pillar to pillar) and sphere-ground (corresponding to pillar head to ground) touching, $F_{VDW}$ can be written[13] as $F_{VDW}$=$-A\sqrt{r_1 r_2}/6D^2$ and $F_{VDW}$=$-Ar/6D^2$, respectively, where $A$= 7×10$^{-20}$ J is the Hamaker constant for the surface (heptadecafluoro-1,1,2,2-tetrahydrodecyl) trimethoxysilane[14] layer, $r_1$ and $r_2$ are the radii of the two pillars, $r$ is the radius of the pillar head sphere, $D$ is the contact distance. Assuming that the two pillars and the pillar tip sphere have the same diameter, then $F_{VDW}$ can be unified as

$$F_{VDW}=-Ar/6D^2 \quad \text{(S5)}$$

For a nanopillar with a diameter of 400 nm, the $F_{VDW}$ as a function of the contact distance $D$ can be calculated and is plotted in Fig. 2f.

For a pillar with a fixed aspect ratio of 10 and is bent at an angle 90°, considering it was touched by another pillar on the tip, the stiction moment caused by the $F_{VDW}$ can be calculated as $F_{VDW}$ ×$x$ and the restoring moment and stiction moment are plotted in Fig. 2i for pillars with different storage modulus. For the pillar head to ground touching, the pillars need to be bent to larger angles (~130° in Fig. 2d). In this case the restoring moment is larger than the 90° touching and similar relation as in Fig. 2i can be obtained.

**7. Control experiment**

To study the influence of the storage modulus on the shape recovery effect, materials with two compositions (mass ratio of AAc, HPPA, and DPEPA 1:3:0.4, 1:1:0.8, referring as soft and stiff compositions respectively) were prepared using the same process as provided in the methods section. Nanopillars were fabricated by using the same printing parameters from these two types of resists. In the programming process, the samples were heated to 20 °C above the glass transition temperature for each composition. The pressure applied were 100 psi, 10 psi for compositions 1~2 respectively. Different pressure was used for different composition to keep the value pressure /storage modulus a constant. After heating the compressed sample to 170 °C, the soft nanopillars did not recover well (Fig. 2g-h) while the stiff nanopillars recovered to the straight state (Fig. 2a-b).

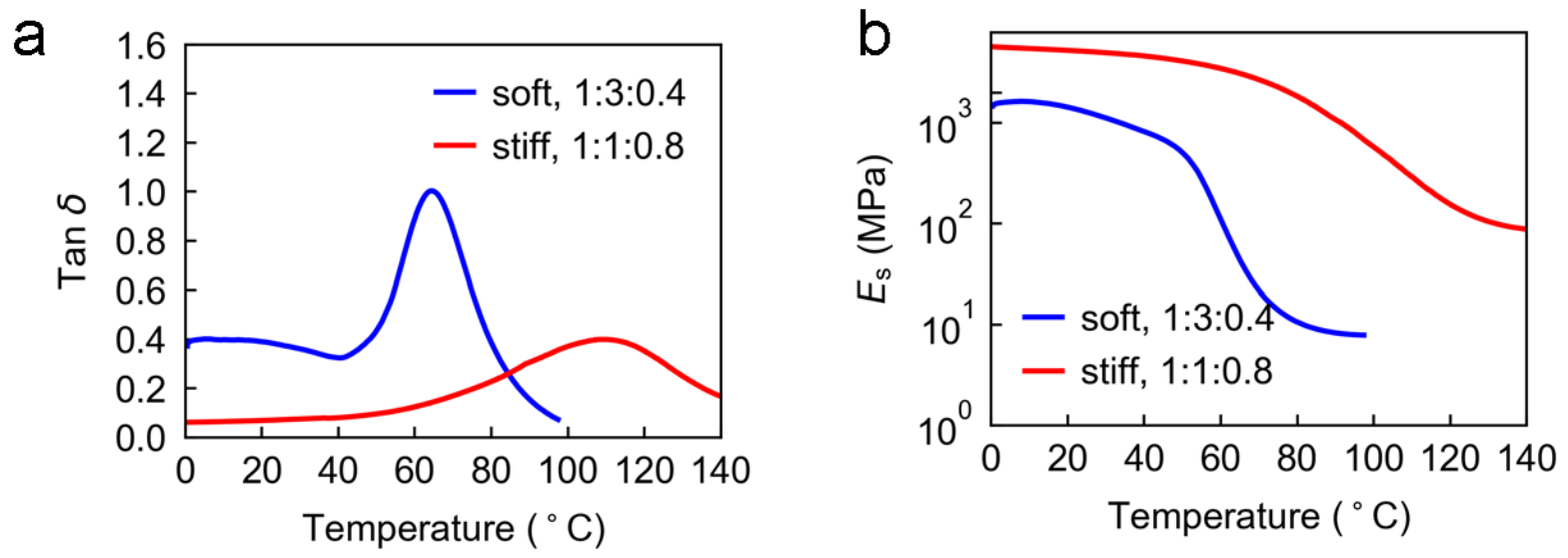

**Supplementary Fig. 5 Influence of modulus on the self-recovery process. a** Tan δ for different compositions (the ratio between AAc, HPPA, and DPEPA are 1:3:0.4, 1:1:0.8, respectively). **b** Storage modulus as a function of temperature for different compositions.

## 8. Self-recovery effect of free-standing grating

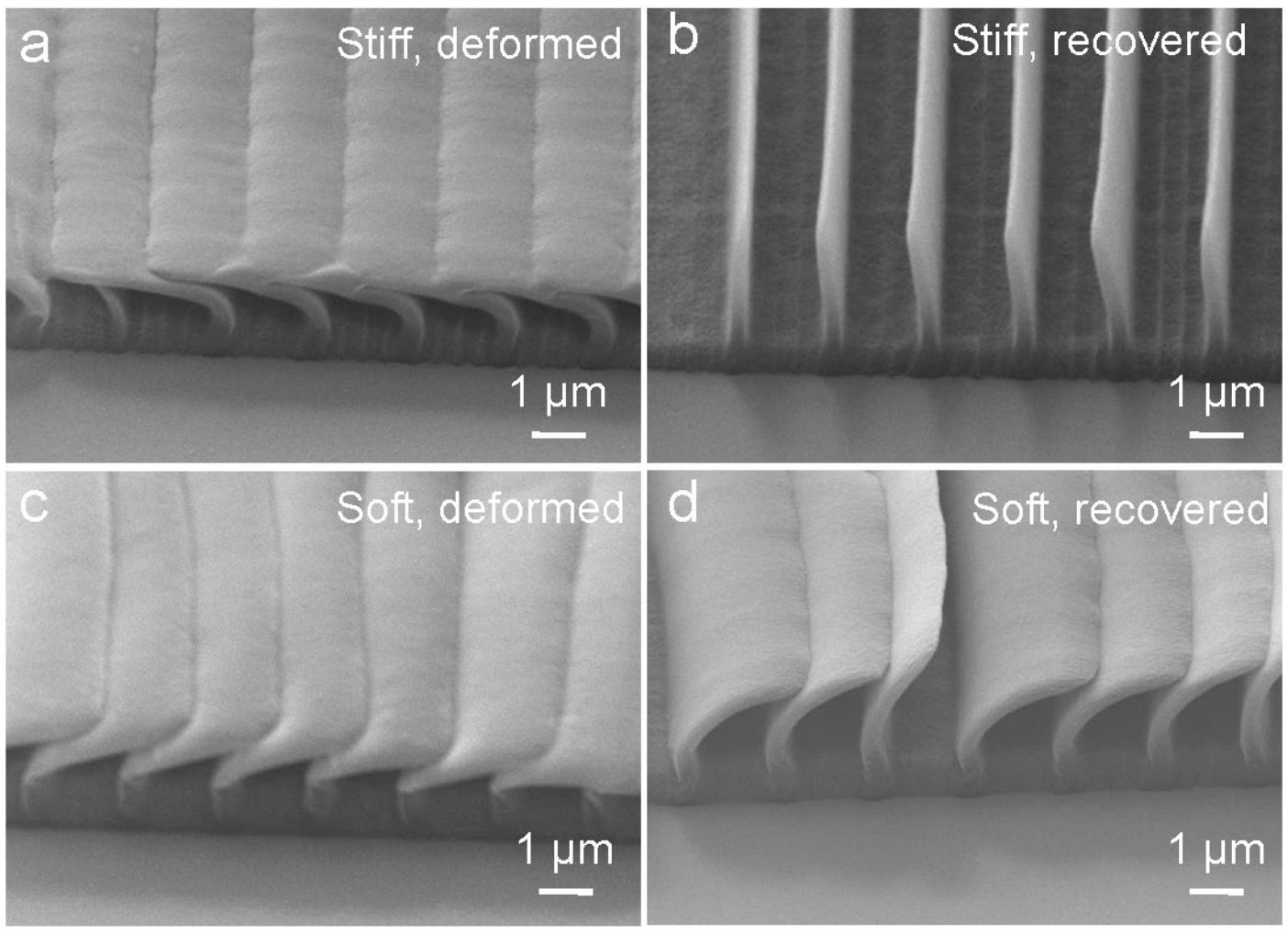

**Supplementary Fig. 6 Self-recovery effect of grating. a-b** Tilted view (45°) SEM images of a grating structure at the deformed and recovered states fabricated by the stiff material ($E_s$=90 MPa). **c-d** Tilted view (45°) SEM images of a grating structure at the deformed and recovered states fabricated by the soft material ($E_s$=10 MPa).

## 9. Study of the influence of pitch and aspect ratio on the recovery ratio

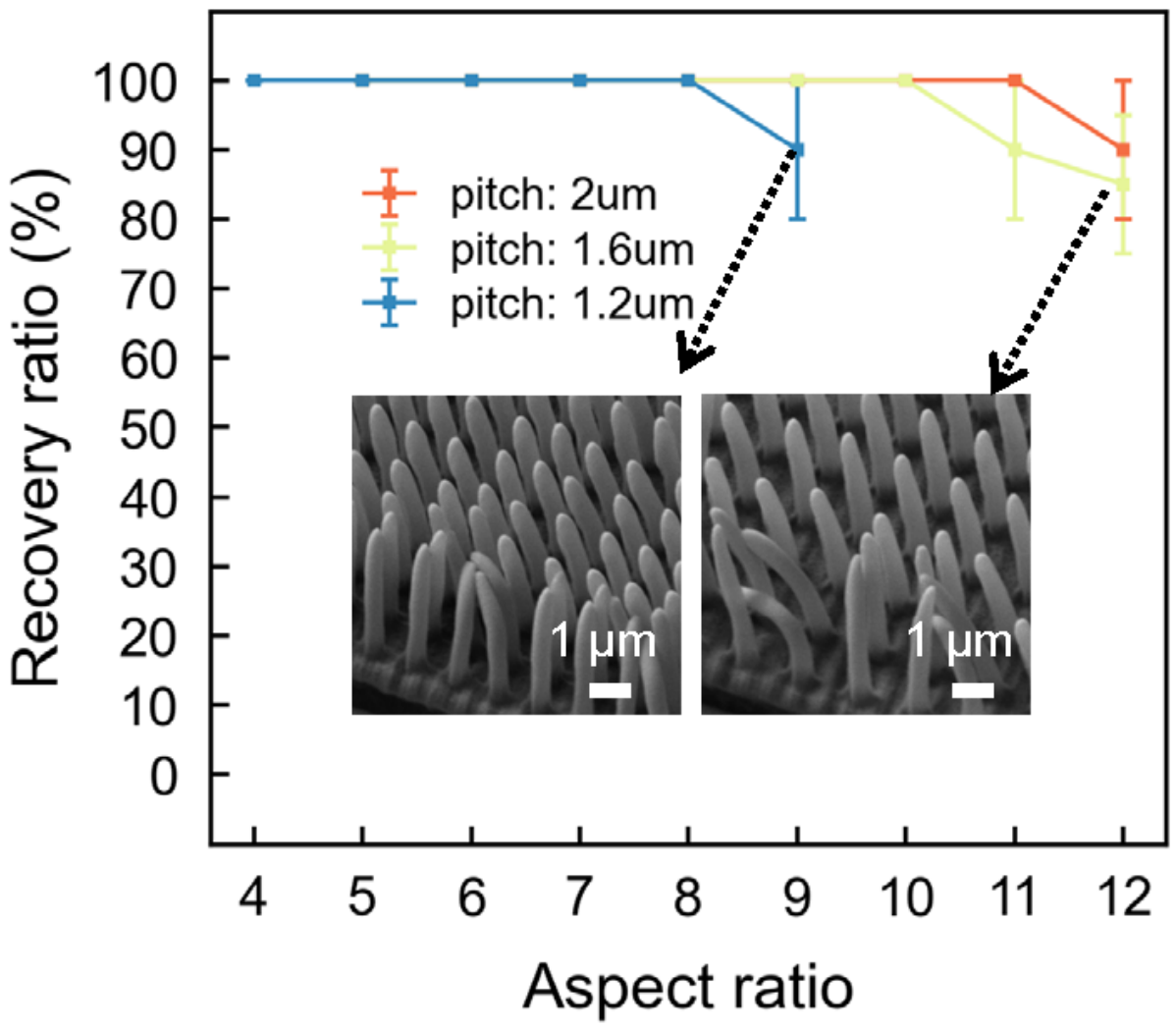

**Supplementary Fig. 7 Influence of aspect ratio and pitch on the shape recovery ratio. Values in the figure represent mean and the error bars represent the standard deviation of the measured values (*n*=5).**

## 10. Collapse during the development and drying process

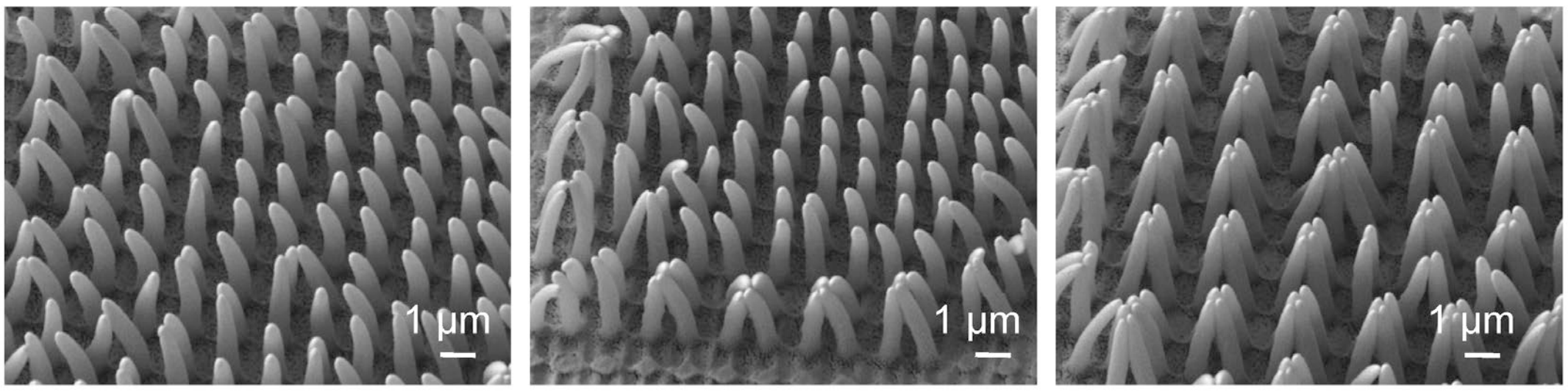

**Supplementary Fig. 8 Tilted view (30°) SEM images indicated the collapse of the as printed nanopillars during the development process for tall nanopillars with 1.2 μm pitch (aspect ratio: 10-12).**

## 11. Structural colour library

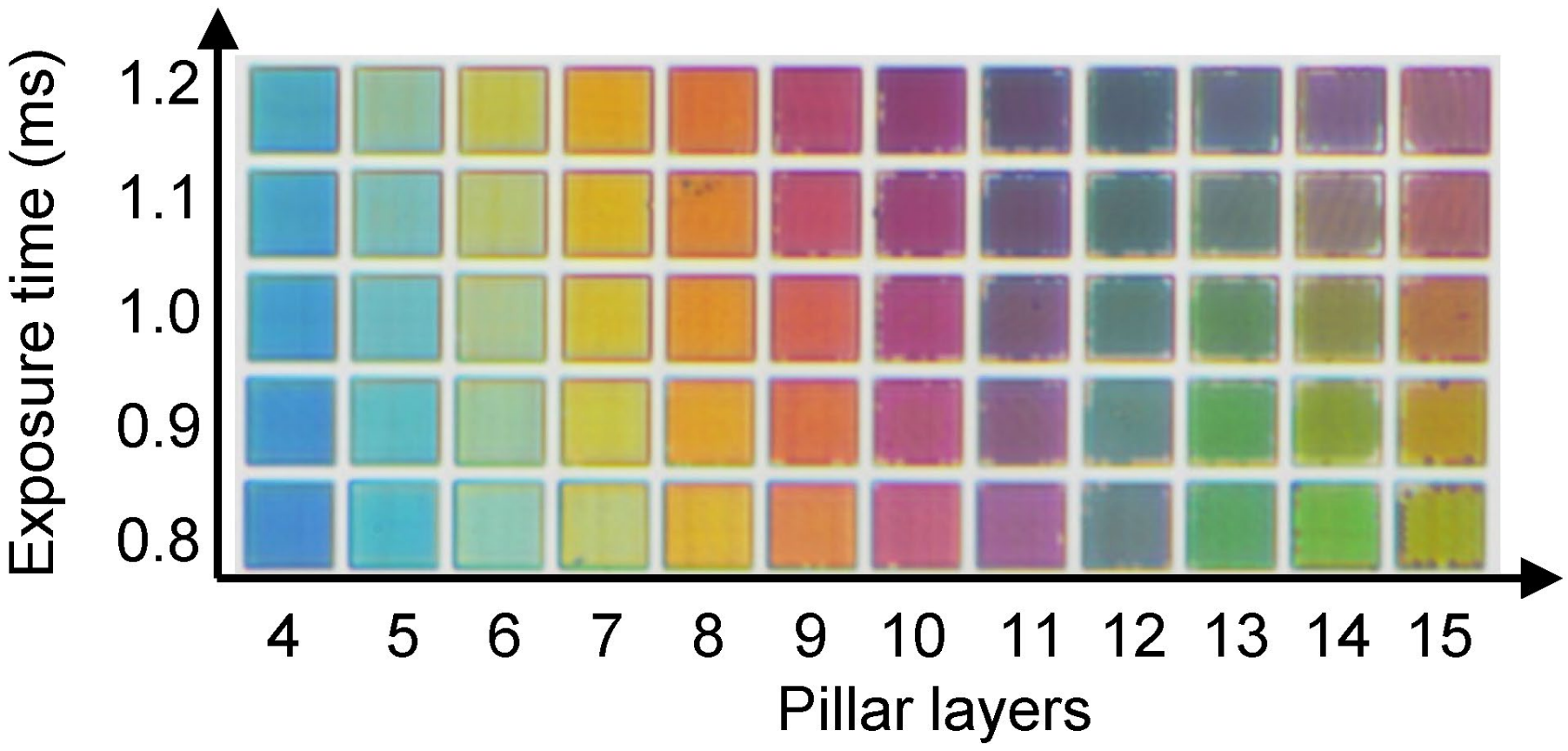

**Supplementary Fig. 9 A structural colour library printed by changing the exposure time and pillar height (pitch=1.2 μm).**

## 12. Finite Element Analysis (FEA)

Finite Element Analysis was used to get the bent shape using the COMSOL Multiphysics using the structural mechanics module. To obtain the material property, a tensile test at the programming temperature (126 °C) was conducted by the DMA Q800 machine. The constitutive equation of the material was obtained by fitting the tensile test data in python codes with the two parameters Mooney-Rivlin hyperelastic model and can be expressed as,

$$\frac{F}{A_0} = 2\left(C_{10} + C_{01}\frac{L_0}{L_0+\Delta L}\right)\left(\frac{L_0+\Delta L}{L_0} - \left(\frac{L_0}{L_0+\Delta L}\right)^2\right) \quad \text{(S6)}$$

Where $A_0$ is the original cross-section area, $L_0$ is the original length, $\Delta L$ is the increase of length, $C_{10}$ = -30.75 MPa and $C_{01}$ = 40.17 MPa are material parameters fitted from the tensile test.

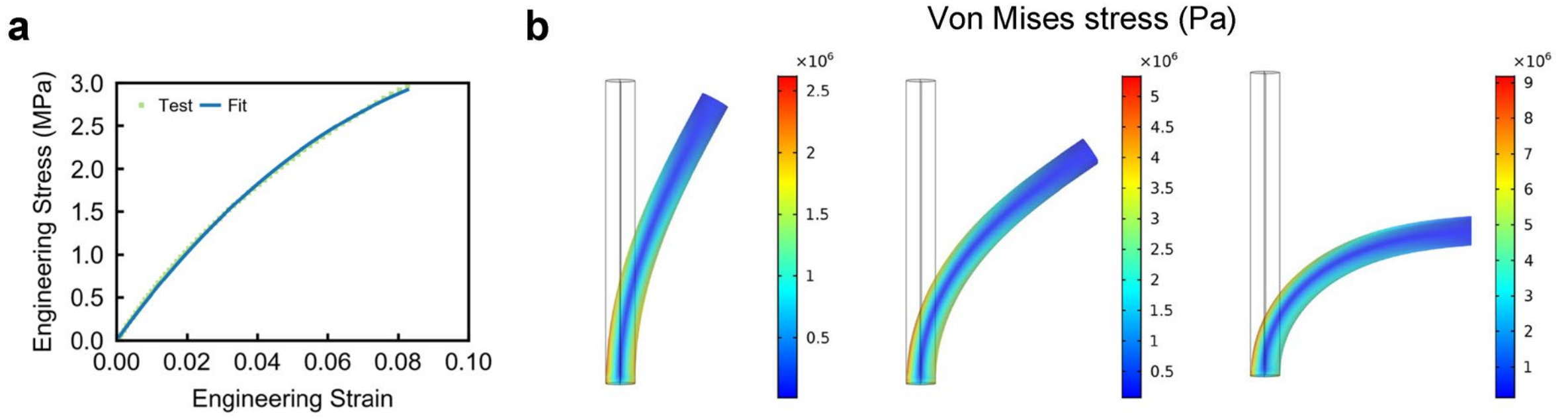

**Supplementary Fig. 10 FEA analysis of a bent nanopillar. a** Tensile test result at the programming temperature (126 °C). **b** Calculated shape of a nanopillar at different bending angles.

## 13. Study of different programmable structural colours

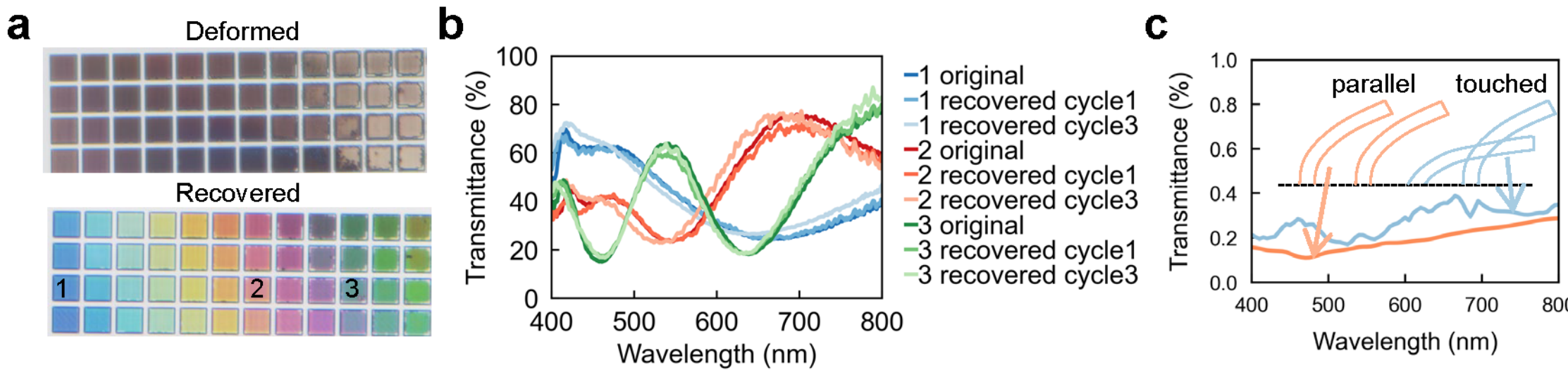

**Supplementary Fig. 11 a** Optical images of deformed and recovered states of different structural colours. **b** Comparison of spectra of different colours (marked as 1-3 in panel a) at the original state and recovered states after different programming cycles. **c** FDTD simulation results for two bent nanopillars with parallel and touched configurations.

## 14. Tuning the amplitude in the Near-Infrared (NIR) light range

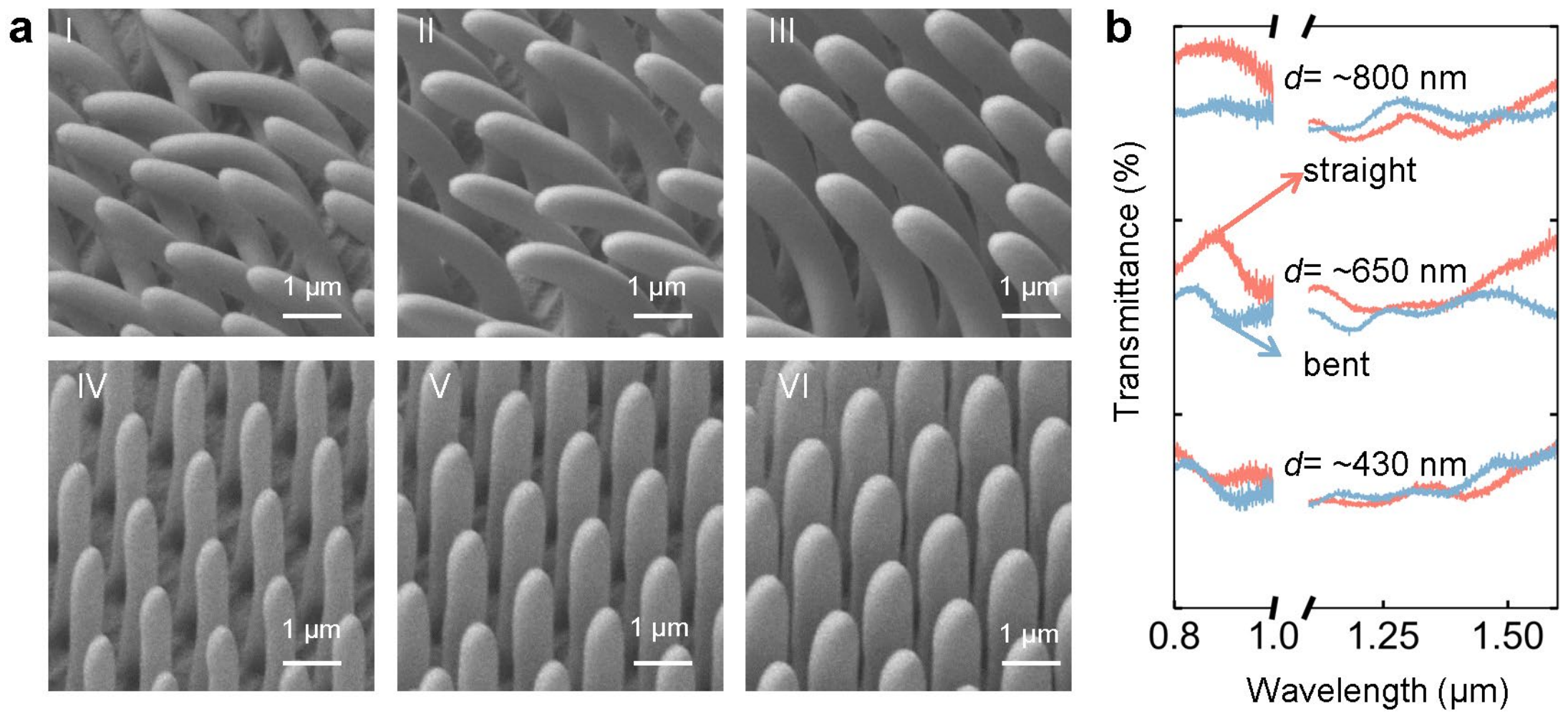

**Supplementary Fig. 12 Tuning the transmittance amplitude in the Near-Infrared (NIR) spectrum. a I-III** Tilted view (45°) SEM images of the bent nanopillars with diameters of ~430 nm, ~650 nm, and ~800 nm respectively and a fixed height of ~5.6 μm. **IV-VI** Tilted view (45°) SEM images of the corresponding nanopillars recovered to the straight state. **b** Measured spectra at the NIR range (0.8-1.0 μm and 1.1-1.6 μm) for nanopillars at the bent and straight states.

## 15. Illustration of different printing modes

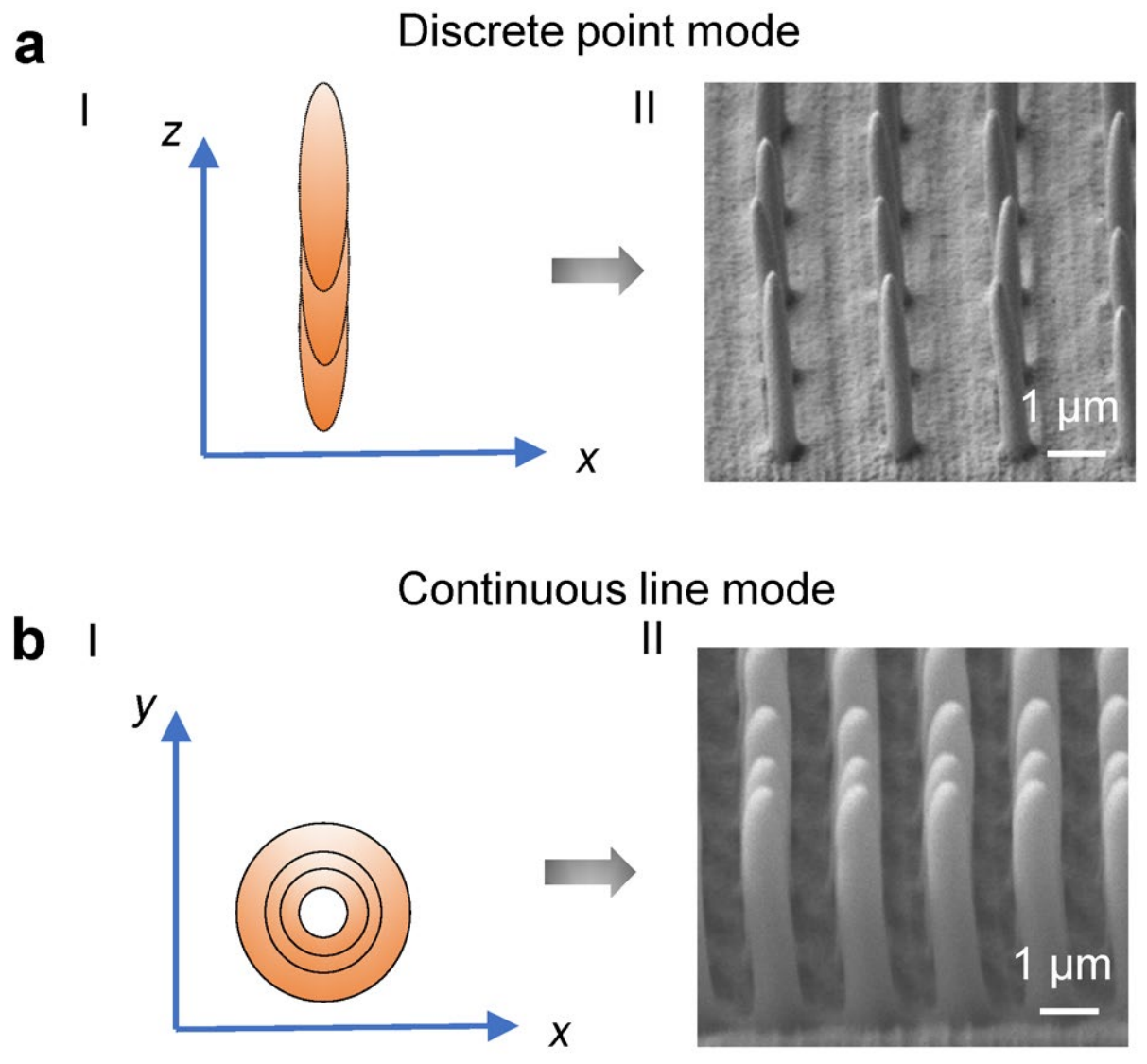

**Supplementary Fig. 13 An illustration of different printing mode. a I** illustration of the discrete point printing mode. **II** Tilted view (45°) SEM image of nanopillars printed by the discrete point mode. **b** I illustration of the continuous line printing mode. **II** Tilted view (45°) SEM image of nanopillars printed by the continuous line mode.

There are two printing modes in the Nanoscribe two photon lithography system, namely discrete point mode (Supplementary Fig. 13aI) and continuous line mode (Supplementary Fig. 13bI). In the discrete point mode, the nanopillars are printed voxel by voxel along z direction by the laser, resulting in smaller diameter of ~400nm (Supplementary Fig. 13aII). In the continuous line mode, the nanopillars are printed with the laser printing concentric circles from inside to outside in the x-y plane for every printing layer, resulting in a wider range of diameter by controlling the number of circles in the x-y plane (Supplementary Fig. 13bII).

### 16. Generation of binary hologram

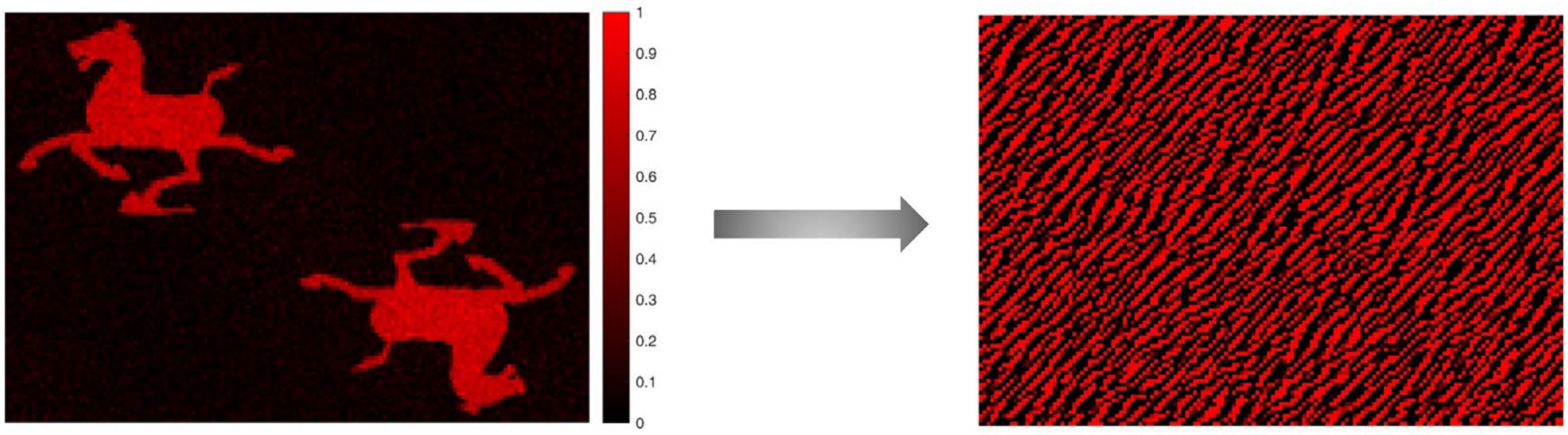

**Supplementary Fig. 14 Design of the binary hologram. a** Input image for hologram generation. **b** binary phase of the corresponding hologram.

The binary phase for hologram is obtained by an adaptive Fourier transform based computer-generated hologram algorithm. We employed the Gerchberg–Saxton algorithm to get the optimized phase with the constraints that only 0 and π values are accepted. To increase the final diffraction efficiency of the hologram, the original image was designed to be centrosymmetric, as seen from the left panel. The right panel is the final phase used, where the red pixels represent phase π and black pixels are 0. The phase map is then converted into the heights of an array of nanopillars, in which nanopillars representing phase π are around 600 nm taller than the ones representing phase 0. The height difference $\Delta h$ is determined by the formula[15] $\Delta h = \Delta\Phi\lambda / 2\pi(n-1)$, where $\Delta\Phi$ is the phase difference, $\lambda$ is the wavelength of the laser, $n$ is the refractive index at the target wavelength. For n~1.5, $\lambda$=635nm, a ~π phase difference can be obtained by printing pillars with ~600 nm difference (3 printing layers).

### 17. Refractive index of the photoresist

The ellipsometry angles $\Delta$ and $\Psi$ were measured by an EP4 ellipsometer (ACCURION, Germany). Then the refractive index $n$ and extinction coefficient $k$ were fitted by the Cauchy dispersion function.

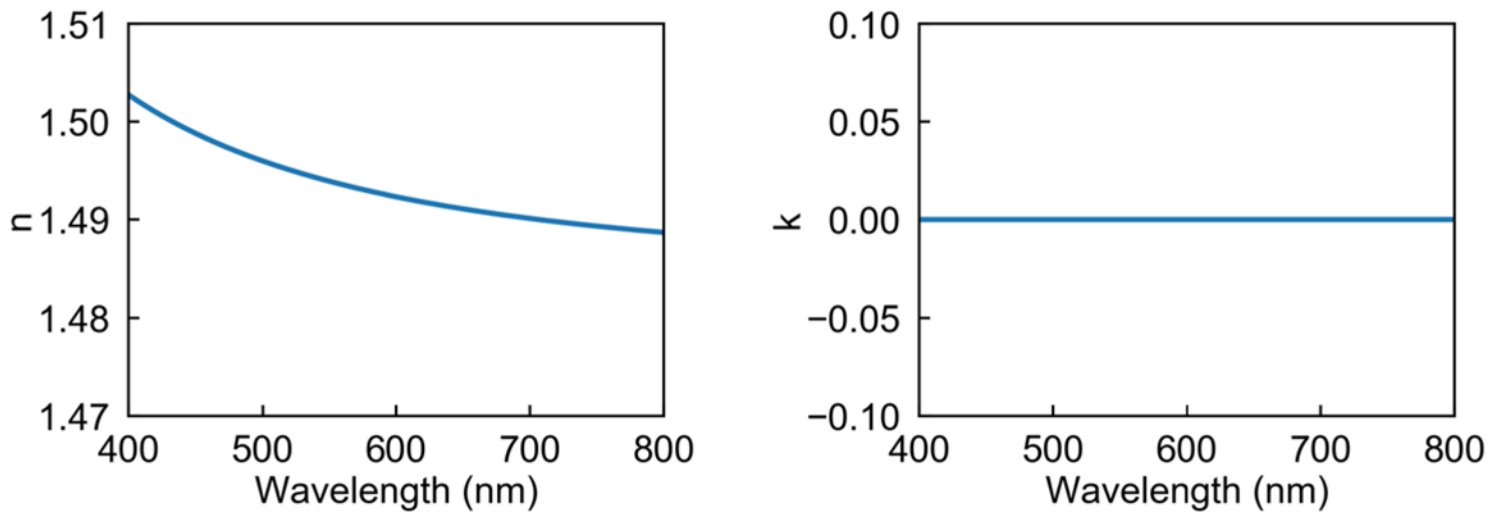

**Supplementary Fig. 15 a** Fitted refractive index *n*. **b** Fitted extinction coefficient *k*.

## Supplementary References

1 Vyatskikh, A. *et al.* Additive manufacturing of 3D nano-architected metals. *Nat. Commun.* **9**, 1-8 (2018).

2 Gall, K. *et al.* Shape memory polymer nanocomposites. *Acta Mater.* **50**, 5115-5126 (2002).

3 Bai, Y., Zhang, X., Wang, Q. & Wang, T. A tough shape memory polymer with triple-shape memory and two-way shape memory properties. *J. Mater. Chem. A* **2**, 4771-4778 (2014).

4 Hearon, K. *et al.* A processable shape memory polymer system for biomedical applications. *Adv. Healthc. Mater.* **4**, 1386-1398 (2015).

5 Zhang, W. *et al.* Structural multi-colour invisible inks with submicron 4D printing of shape memory polymers. *Nat. Commun.* **12**, 1-8 (2021).

6 Zhang, B. *et al.* Self-healing four-dimensional printing with an ultraviolet curable double-network shape memory polymer system. *ACS Appl. Mater. Interfaces* **11**, 10328-10336 (2019).

7 Yang, C. *et al.* 4D printing reconfigurable, deployable and mechanically tunable metamaterials. *Mater. Horiz.* **6**, 1244-1250 (2019).

8 Chen, C. M. & Yang, S. Directed water shedding on high-aspect-ratio shape memory polymer micropillar arrays. *Adv. Mater.* **26**, 1283-1288 (2014).

9 Kuang, X. *et al.* 3D printing of highly stretchable, shape-memory, and self-healing elastomer toward novel 4D printing. *ACS Appl. Mater. Interfaces* **10**, 7381-7388 (2018).

10 Zarek, M. *et al.* 3D printing of shape memory polymers for flexible electronic devices. *Adv. Mater.* **28**, 4449-4454 (2016).

11 Zhang, B. *et al.* Mechanically Robust and UV-Curable Shape-Memory Polymers for Digital Light Processing Based 4D Printing. *Adv. Mater.*, 2101298 (2021).

12 Landau, L. D. & Lifshitz, E. M. *Theory of elasticity*. Vol. 7 (Pergamon Press, Oxford New York, 1986).

13 Israelachvili, J. N. *Intermolecular and surface forces*. (Academic press, 2015).

14 Ohnishi, S., Yaminsky, V. V. & Christenson, H. K. Measurements of the force between fluorocarbon monolayer surfaces in air and water. *Langmuir* **16**, 8360-8367 (2000).

15 Wang, H., Wang, H., Zhang, W. & Yang, J. K. Toward near-perfect diffractive optical elements via nanoscale 3D printing. *ACS Nano* **14**, 10452-10461 (2020).